\documentclass[11pt]{article} %
\usepackage{graphicx}
\usepackage{fancyhdr}
\usepackage{makeidx}
\usepackage{amssymb,amsmath,pdfpages}
\usepackage{physics}
\usepackage{gensymb}
\usepackage{upquote}
\usepackage{ulem}
\usepackage{xcolor}
\usepackage{cite}
\usepackage{chemgreek}
\usepackage[utf8]{inputenc}
\usepackage[nottoc]{tocbibind}
\usepackage{setspace}
\usepackage{enumerate}
\DeclareUnicodeCharacter{0308}{HERE!HERE!}
\DeclareUnicodeCharacter{030C}{HERE!HERE!}
\newcommand{\bd}{\begin{document}}
	\newcommand{\ed}{\end{document}}
\newcommand{\bc}{\begin{center}}
	\newcommand{\ec}{\end{center}}
\newcommand{\vs}{\vspace}
\newcommand{\hs}{\hspace}
\newcommand{\beq}{\begin{equation}}
\newcommand{\eeq}{\end{equation}}
\newcommand{\beqs}{\begin{eqn*}}
	\newcommand{\eeqs}{\end{eqn*}}
\newcommand{\bq}{\begin{quote}}
	\newcommand{\eq}{\end{quote}}
\newcommand{\lb}{\linebreak}
\newcommand{\mb}{\makebox}
\newcommand{\fb}{\framebox}
\newcommand{\mc}{\multicolumn}
\newcommand{\ben}{\begin{enumerate}}
	\newcommand{\een}{\end{enumerate}}
\newcommand{\bit}{\begin{itemize}}
	\newcommand{\eit}{\end{itemize}}
\newcommand{\ov}{\overline}
\newcommand{\un}{\underline}
\newcommand{\lt}{\left}
\newcommand{\rt}{\right}
\newcommand{\ba}{\begin{array}}
	\newcommand{\ea}{\end{array}}
\newcommand{\beqa}{\begin{eqnarray}}
\newcommand{\eeqa}{\end{eqnarray}}
\newcommand{\beqas}{\begin{eqnarray*}}
	\newcommand{\eeqas}{\end{eqnarray*}}
\newcommand{\bfg}{\begin{figure}}
	\newcommand{\efg}{\end{figure}}
\newcommand{\pad}{\partial}
\newcommand{\nn}{\nonumber}
\newcommand{\la}{\leftarrow}
\newcommand{\ra}{\rightarrow}
\newcommand{\lgla}{\longleftarrow}
\newcommand{\lgra}{\longrightarrow}
\newcommand{\La}{\Leftarrow}
\newcommand{\Ra}{\Rightarrow}
\newcommand{\Lra}{\Leftrightarrow}
\newcommand{\Lgla}{\Longleftarrow}
\newcommand{\Lgra}{\Longrightarrow}
\renewcommand{\a}{\alpha}
\renewcommand{\b}{\beta}
\newcommand{\g}{\gamma}
\newcommand{\G}{\Gamma}
\renewcommand{\d}{\delta}
\newcommand{\D}{\Delta}
\newcommand{\e}{\epsilon}
\newcommand{\eps}{\epsilon}
\newcommand{\s}{\sigma}
\renewcommand{\l}{\lamda}
\newcommand{\m}{\mu}
\newcommand{\n}{\nu}
\renewcommand{\S}{\Sigma}
\newcommand{\p}{\pi}
\newcommand{\om}{\omega}
\newcommand{\Om}{\Omega}
\newcommand{\tri}{\triangle}
\newcommand{\ti}{\times}
\newcommand{\f}{\frac}
\newcommand{\ds}{\displaystyle}
\newcommand{\bm}[1]{\mb{{\boldmath $#1$}}}
\newcommand{\alter}[2]{\lt\{ \ba{ll}#1 \\ #2 \ea \rt.}
\newcommand{\alt}[4]{\lt\{ \ba{ll}#1 & \mb{if \, \,}#2 \\ #3 & \mb{}#4 \ea
	\rt.}
\newcommand{\altn}[4]{\lt\{ \ba{rl}#1 & \mb{if \, \,}#2 \\ #3 & \mb{}#4 \ea
	\rt.}
\newcommand{\altif}[4]{\lt\{ \ba{ll}#1 & \mb{if \, \,}#2 \\ #3 &
	\mb{if \, \,}#4 \ea \rt.}
\newcommand{\altnif}[4]{\lt\{ \ba{rl}#1 & \mb{if \, \,}#2 \\ #3 &
	\mb{if \, \,}#4 \ea \rt.}
\newcounter{algc}
\newcounter{romc}
\newcounter{Alphc}
\newcommand{\bl}{\begin{list}{{\it Step} ~\arabic{algc}~:} {\usecounter{algc}
			\setlength{\topsep}{0pt} \setlength{\itemsep}{0pt}}}
	\newcommand{\el}{\end{list}}
\newcommand{\blr}{\begin{list}{~\roman{romc}~:} {\usecounter{romc}
			\setlength{\topsep}{0pt} \setlength{\itemsep}{0pt}}}
	\newcommand{\elr}{\end{list}}
\newcommand{\bla}{\begin{list}{~\Alph{Alphc}~:} {\usecounter{Alphc}
			\setlength{\topsep}{0pt} \setlength{\itemsep}{0pt}}}
	\newcommand{\ela}{\end{list}}
\newcommand{\tsup}{\textsuperscript}
\newcommand{\tsub}{\textsubscript}

\newtheorem{theorem}{Theorem}
\begin{document}
 \title{Trion-trion annihilation in monolayer WS\tsub2}
 \author{Suman Chatterjee$^{1}$, Garima Gupta$^{1}$, Sarthak Das$^{1}$, \\Kenji Watanabe$^2$, Takashi Taniguchi$^3$ and Kausik Majumdar$^{1*}$\\
	$^1$Department of Electrical Communication Engineering, \\Indian Institute of Science, Bangalore 560012, India\\
	$^2$Research Center for Functional Materials,\\ National Institute for Materials Science, 1-1 Namiki, Tsukuba 305-044, Japan\\
	$^3$International Center for Materials Nanoarchitectonics,\\ National Institute for Materials Science, 1-1 Namiki, Tsukuba 305-044, Japan\\
	%$^{\dagger}$These authors contributed equally,\\
	$^*$Corresponding author, email: kausikm@iisc.ac.in}
%\date{}
\maketitle
\begin{abstract}
 Strong Coulomb interaction in monolayer transition metal dichalcogenides can facilitate nontrivial many-body effects among excitonic complexes. Many-body effects like exciton-exciton annihilation (EEA) have been widely explored in this material system. However, a similar effect for charged excitons (or trions), that is, trion-trion annihilation (TTA), is expected to be relatively suppressed due to repulsive like-charges, and has not been hitherto observed in such layered semiconductors. By a gate-dependent tuning of the spectral overlap between the trion and the charged biexciton through an ``anti-crossing''-like behaviour in monolayer WS$_2$, here we present an experimental observation of an anomalous suppression of the trion emission intensity with an increase in gate voltage. The results strongly correlate with time-resolved measurements, and are inferred as a direct evidence of a nontrivial TTA resulting from non-radiative Auger recombination of a bright trion, and the corresponding energy resonantly promoting a dark trion to a charged biexciton state. The extracted Auger coefficient for the process is found to be tunable ten-fold through a gate-dependent tuning of the spectral overlap.
 \end{abstract}

\newpage
\section*{Introduction}
Semiconducting monolayers of transition metal dichalcogenides (TMDCs) exhibit strongly bound excitons and other higher-order excitonic complexes \cite{mak2013tightly,ross2013electrical,singh2016trion,you2015observation,ye2018efficient,robert2017fine,chen2018coulomb,chatterjee2021probing,paur2019electroluminescence}. Under high excitation density, large Coulomb interaction in these sub-nanometer-thick monolayers can lead to a strong interaction among these excitonic complexes \cite{erkensten2021exciton,leon2019hot,linardy2020harnessing,chow2020monolayer,cunningham2016auger,singh2014coherent,singh2016trion}. Among several many-body effects, exciton-exciton annihilation (EEA), as determined by the Auger interaction between two excitons, has particularly drawn a lot of attention from the researchers \cite{erkensten2021exciton,leon2019hot,chow2020monolayer,hoshi2017suppression,sun2014observation,lee2022enhanced}. In a typical Auger process in bulk semiconductors (like Si or Ge), the recombination energy of two quasiparticles (electron and hole) is taken up by a second electron, and is promoted to a higher energy state \cite{salehzadeh2014exciton,yablonovitch1986auger,beattie1959auger}. The excitonic Auger process involves two excitons where one recombines non-radiatively, providing the recombination energy to a second exciton. This energy usually ionizes the receiving exciton \cite{leon2019hot,erkensten2021exciton,hoshi2017suppression}. Since these processes directly affect the quantum efficiency of optical devices \cite{salehzadeh2014exciton,kuroda2020dark}, a clear understanding of such processes is of utmost importance. On the other hand, Auger interaction can also be used to produce useful photocurrent through the generation of hot carriers \cite{chow2020monolayer,linardy2020harnessing}.

One of the essential advantages of monolayer TMDCs lies in the gate tunability of the optical response. Under gating, the neutral exciton can easily transfer its oscillator strength to a charged exciton (or trion). Similar many-body effects like trion-trion annihilation (TTA) should be intriguing since, on the one hand, the relatively larger Bohr radius of trions may enhance the Auger process. In contrast, on the other hand, trion-trion repulsion due to like-charges would suppress such interaction. While Auger processes in higher-order excitonic complexes have been demonstrated in quantum dots \cite{park2014auger}, no such TTA effect has been reported to date in monolayer TMDCs. In this work, we present an experimental observation of the same in gated monolayer WS\tsub2.

\section*{Results and Discussions}
\begin{figure}[!hbt]
	\centering
	\vs{-0.1in}
	\hs{-0.0in}
	\includegraphics[scale=0.5]{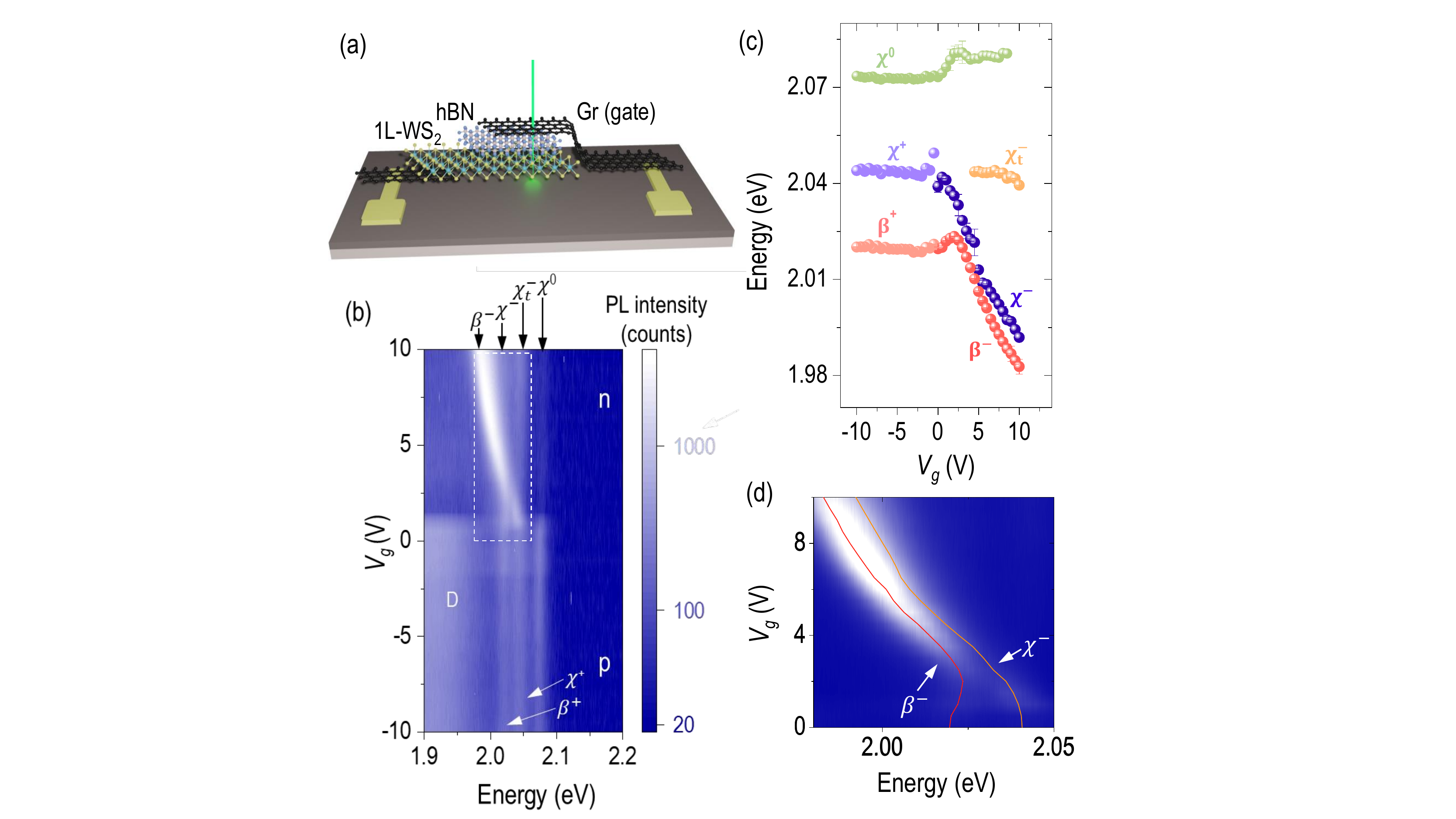}
	\vspace{-0.25in}
	\caption{\textbf{Gate voltage-dependent modulation of excitonic complexes in WS$_2$.} (a) Schematic of 1L WS\tsub2/hBN/few-layer graphene device. A gate voltage ($V_g$) is applied between the two contact pads.  (b) Color plot of $V_g$ dependent PL intensity (in log scale) and the spectral position variation of the neutral exciton ($\chi^0$), positive and negative trions ($\chi^{\pm}$), Inter-valley trion( $\chi_t^-$), and positive and negative charged biexcitons ($\beta^{\pm}$). p-doping and n-doping regions are marked. D stands for the broad defect bound exciton emission peak. (c) Fitted peak positions of all the relevant peaks plotted as a function of $V_g$. The error bars obtained from multiple fits are indicated. (d) Zoomed in portion (marked by white dashed box) of the color plot (a), plotted in linear scale, depicting (red and yellow guide to eye lines represent trion and biexciton respectively) ``anti-crossing''-like behaviour between $\chi^-$ and $\beta^-$.} \label{fig:pl peak position}
\end{figure}

%Sample preparation and stack measurement
We prepare a stack of 1L-WS\tsub2/hBN/few-layer graphene on a heavily-doped Si wafer covered with 285 nm SiO\tsub2 using the dry transfer method. The stack is then annealed in high vacuum at $200^\circ$C for $3$ hours (see \textbf{Methods} section for sample preparation). A gate voltage ($V_g$) is applied to the top graphene electrode while connecting 1L-WS\tsub2 to a grounded few-layer graphene contact. The device configuration is schematically depicted in Figure \ref{fig:pl peak position}a. \textcolor{black}{According to previous reports \cite{hoshi2017suppression,lee2022enhanced}, 1L-WS\tsub2 on Si/SiO\tsub2 exhibits enhanced Auger effect due to localization of excitonic species. In this work, we thus intentionally keep the 1L-WS\tsub2 flake directly on the Si/SiO\tsub2 substrate (only covered by top hBN, see Figure \ref{fig:pl peak position}a) to achieve an overall higher Auger rate constant \cite{hoshi2017suppression}.} All measurements are performed in an optical cryostat at $T = 5$ K unless otherwise mentioned. We excite the sample with a pulsed laser (531 nm) and collect both steady-state photoluminescence (PL) and time-resolved photoluminescence (TRPL) \emph {in situ} (see Supplemental Material \cite{Supp.Mater.} Figure 1 for setup schematic) at different $V_g$.

The spectral evolution of different excitonic peaks with $V_g$ is depicted in a color plot in Figure \ref{fig:pl peak position}b (all the individual spectra are shown in Supplemental Material \cite{Supp.Mater.} Figure 2). Figure \ref{fig:pl peak position}c shows the respective fitted peak positions. We explain the origin of such non-monotonic feature from a combination of three competing effects: (1) $V_g$ induces charge carriers in the monolayer, the resulting screening weakens the Coulomb interaction causing a suppression of the binding energy of the excitonic complexes, which in turn results in a blue shift in the emission energy \cite{wang2017probing,li2021refractive,van2017marrying}; (2) the vertical gate field induces a spatial separation between the electrons and holes in the excitonic complexes, leading to a suppression in the binding energy, and hence a blue shift in the emission energy \cite{abraham2021anomalous}; and (3) doping induced Pauli blocking, leading to a redshift in emission \cite{kallatt2019interlayer,mak2013tightly,huard2000bound,stebe1989ground}. The last effect is the strongest one in our sample, as explained later in more detail.

The p-doping side ($V_g < 0$ V) shows three distinct features (Figure \ref{fig:pl peak position}b and c), namely, neutral bright exciton ($\chi^0$), positive trion ($\chi^+$), and positively charged biexciton ($\beta^+$). Due to the relative positions of the Fermi levels between 1L-WS\tsub2 and graphene, it is more difficult to inject holes than electrons \cite{murali2021accurate}. Thus it is challenging to dope the 1L-WS\tsub2 flake strongly p-type, as suggested by the sustained $\chi^0$ peak intensity even at high negative $V_g$ (Figure \ref{fig:pl peak position}b). The inability to strongly dope the WS\tsub2 film p-type leads to a negligible doping-induced spectral-shift of the peaks for $V_g<0$ [as effects (1) and (3) are negligible]. In contrast, we observe a small blue shift (about 2-4 meV) for all the three peaks, shown in Supplementary Material \cite{Supp.Mater.} Figure 3. In the absence of a significant p-doping, such a blue shift provides an estimate for the reduction in the binding energy due to gate-field-induced spatial separation of the electron and hole in an excitonic complex \cite{abraham2021anomalous}.

The spectral features are rich in the n-doping side ($V_g >0$ V). With an increase in doping density, the $\chi^0$ peak gradually fades out due to a transfer of its oscillator strength to the negatively charged trion ($\chi^-$) \cite{mak2013tightly,das2020highly,rana2021many,wagner2020autoionization,ross2013electrical} and charged biexciton ($\beta^-$), as indicated in Figure \ref{fig:pl peak position}b. The biexcitonic nature of the $\beta^-$ peak is identified through a super-linear power law \cite{chatterjee2021probing,barbone2018charge,you2015observation} (see Supplementary Material \cite{Supp.Mater.} Figure 4). While all the three peaks show a slight blue shift at low positive $V_g$, with an increase in voltage, both the $\chi^-$ (for $V_g >1$ V) and $\beta^-$ (for $V_g >2.5$ V) peaks exhibit a strong redshift (zoomed in Figure \ref{fig:pl peak position}d). On the other hand, the $\chi^0$ peak position remains nearly unaltered after an initial blue shift (Figure \ref{fig:pl peak position}c).

After $V_g > 4.5$ V, another distinct trion peak emerges around $2.04$ eV - in a spectral position similar to that of $\chi^+$ (Figure \ref{fig:pl peak position}c). We attribute this to the formation of negative intervalley trion ($\chi_t^-$) from the top of the spin-split conduction bands at high $V_g$, as shown in Supplementary Material \cite{Supp.Mater.} Figure 5. It appears after a certain threshold value of $V_g$ (4.5 V) due to a lower filling of the top conduction bands. This peak shows negligible redshift due to weak Pauli blocking at higher energy unless $V_g$ is very high ($> 8$ V).

\begin{figure}[!hbt]
	\centering
	\vs{-0.1in}
	\hs{-0in}
   \includegraphics[scale=0.5]{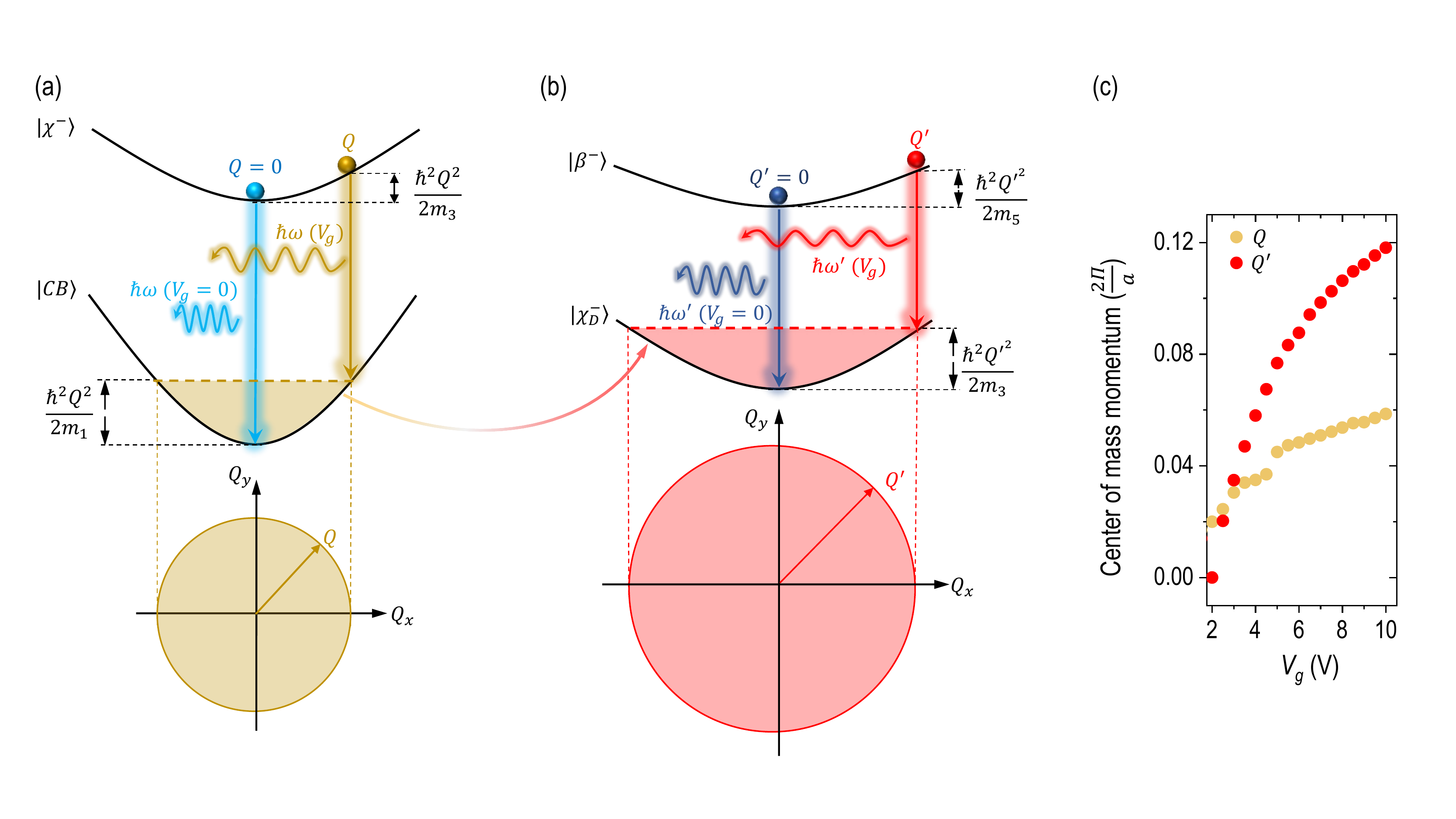}
	\vspace{-0.5in}
	\caption{\textbf{Mechanism for gate-dependent redshift in emission from trion and charged biexciton with non-zero center of mass momenta.} (a) The vertical transition from the negative trion state ($\ket{\chi^-}$) to the conduction band ($\ket{CB}$), emitting one photon, is shown at two different $V_g$ values. At $V_g = 0$ V, $\ket{CB}$ is almost empty, and transition from the bottom of the $\ket{\chi^-}$ to $\ket{CB}$ ($Q = 0$) is possible (shown by sky-blue arrow). With increasing n-doping ($V_g> 0$ V), lower energy $\ket{CB}$ states are blocked, and the vertical transition occurs at a higher $Q$ value (shown by yellow arrow). In-plane ($Q_x$-$Q_y$) slice shows filling of bands up to a radius of $Q$ by yellow shaded area (bottom panel). The dissimilar curvatures of the two bands result in a redshift in emission at larger filling. (b) The vertical transition from the negatively charged biexciton state ($\ket{\beta^-}$) to dark trion ($\ket{\chi_D^-}$) state, emitting one photon at two different $V_g$ values. $V_g = 0$ V transition occurs at $Q^\prime = 0$ (shown by dark blue arrow). At finite $V_g$, the $\ket{\chi_D^-}$ state is blocked up to $Q^\prime$, pushing the vertical transition to occur at higher $Q^\prime$ (shown by red arrow). In-plane ($Q_x$-$Q_y$) slice area, shaded in red, indicates a larger filling compared to (a) (bottom panel). Loss of electrons from conduction band in order to form $\chi_D^-$ is represented by a shaded arrow from $\ket{CB}$ to $\ket{\chi_D^-}$. (c) COM momenta of the emitting $\chi^-$ and $\beta^-$ species plotted as a function of $V_g$ indicating a cross over at higher $V_g$.} \label{fig:Q values}
\end{figure}

As indicated in Figure \ref{fig:Q values}a-b, the emission energy of the $\chi^-$ ($\beta^-$) peak corresponds to the transition from the trion (charged biexciton) band to the electron (trion) band. Consequently, the trion or the charged biexciton with a large center of mass (COM) momentum can emit a photon. This is in stark contrast with the case of a neutral exciton, where the conservation of linear momentum during radiative recombination allows only near-zero momentum states (within the light cone\cite{gupta2019fundamental}) to emit a photon. As shown in Figure \ref{fig:Q values}a, for the trion, as $V_g$ increases, smaller momentum states are filled in the conduction band, and hence the trions with small momentum are not allowed to recombine due to Pauli blocking of the final state. Hence, as $V_g$ increases, the allowed trion emission moves to a higher momentum. This effect, in turn, results in a reduction in the trion emission energy due to the different curvatures of the two participating bands. The corresponding redshift $\Delta(V_g) = \hbar\omega(V_g=0) - \hbar\omega(V_g)$ can be directly mapped to the momentum ($\hbar Q$) of the emitting trion following the equation:
\beq\label{eq:tx}
\Delta(V_g) = \frac{\hbar^2Q^2}{2}\bigg(\frac{1}{m_1} - \frac{1}{m_3}\bigg)
\eeq
In the above equation, we neglect any minor blue shift due to screening and field effect. We also assume that there is negligible filling in the conduction band (CB) around $V_g=0$, thus, $V_g=0$ corresponds to transitions at $Q=0$ (shown in Figure \ref{fig:Q values}a). $m_1$ and $m_3$ are the conduction band effective mass and the center of mass of the trion, respectively.

Similarly, the charged biexciton emission moves to a larger momentum with an increase in $V_g$ due to Pauli blocking for the final trion state, and results in a similar red shift due to the difference in curvature between the charged biexciton and trion bands. The COM momentum $\hbar Q^\prime$ of the emitting charged biexciton is related to the corresponding red shift [$\Delta^\prime(V_g) = \hbar\omega^\prime(V_g=0) - \hbar\omega^\prime(V_g)$] as
\beq\label{eq:bx}
\Delta^\prime(V_g) = \frac{\hbar^2{Q^\prime}^2}{2}\bigg(\frac{1}{m_3} - \frac{1}{m_5}\bigg)
\eeq
where $m_5$ is the mass of the charged biexciton.

Using equations \ref{eq:tx} and \ref{eq:bx}, in Figure \ref{fig:Q values}c we plot the extracted $Q$ and $Q^\prime$ from the measured redshifts in Figure \ref{fig:pl peak position}c as a function of $V_g$. One striking observation is that, for $V_g>2.5$ V, $Q^\prime$ is larger than $Q$. The large separation between $Q$ and $Q^\prime$ suggests that the charged biexciton emission occurs at a larger momentum value compared with the emitting bright trion, indicating that the final state of the charged biexciton emission must be filled up to a momentum $\hbar Q^\prime$ ($>\hbar Q$). This, in turn, suggests that the final state of charged biexciton emission is a dark trion (${\chi_D}^-$), since the states in the bright trion band are empty for momentum $>\hbar Q$, and thus should have allowed transitions for momentum just above $\hbar Q$ (and hence, $Q^\prime$ should have been close to $Q$). Hence, it is instructive to consider that the charged biexciton is formed from a bright exciton and a dark trion ($\chi_D^-$). The long-lived nature of the $\chi_D^-$ \cite{volmer2017intervalley,liu2019gate,li2019direct} keeps the band filled up to a momentum $\hbar Q^\prime (>\hbar Q)$. Such charged biexciton configuration with $\chi_D^-$ and $\chi^0$ is in agreement with previous reports \cite{zinkiewicz2021excitonic,barbone2018charge}.

The zoomed-in version of the PL color plot in Figure \ref{fig:pl peak position}d, (guide to eye lines are drawn in yellow and red to represent $\chi^-$ and $\beta^-$ peak position) exhibits a striking ``anti-crossing''-like feature: the two peaks initially come close to each other, followed by a repulsion. Due to a large difference in the curvature between the initial and final states of the $\chi^-$ emission, the spectral red shift of this peak is initially much larger  compared with the $\beta^-$ peak (see Figure \ref{fig:Q values}a-b). This effect, coupled with a gate field induced blue shift of the $\beta^-$ peak causes these two peaks to come closer at lower $V_g$. However, when $V_g$ increases, the conduction band filling slows down as the electrons start forming the long-lived dark trions (shown by a shaded curved arrow between Figure \ref{fig:Q values}a and b). This, on one hand, slows down the spectral movement of the $\chi^-$ peak, while on the other hand, enhances the spectral red shift of the $\beta^-$ emission peak - forcing a spectral repulsion between the two peaks. This results in a $V_g$ dependent tunable spectral overlap between the $\chi^-$ and $\beta^-$ peaks.

\begin{figure}[!hbt]
	\centering
	\vs{-0.1in}
	\hs{-0in}
   \includegraphics[scale=0.5]{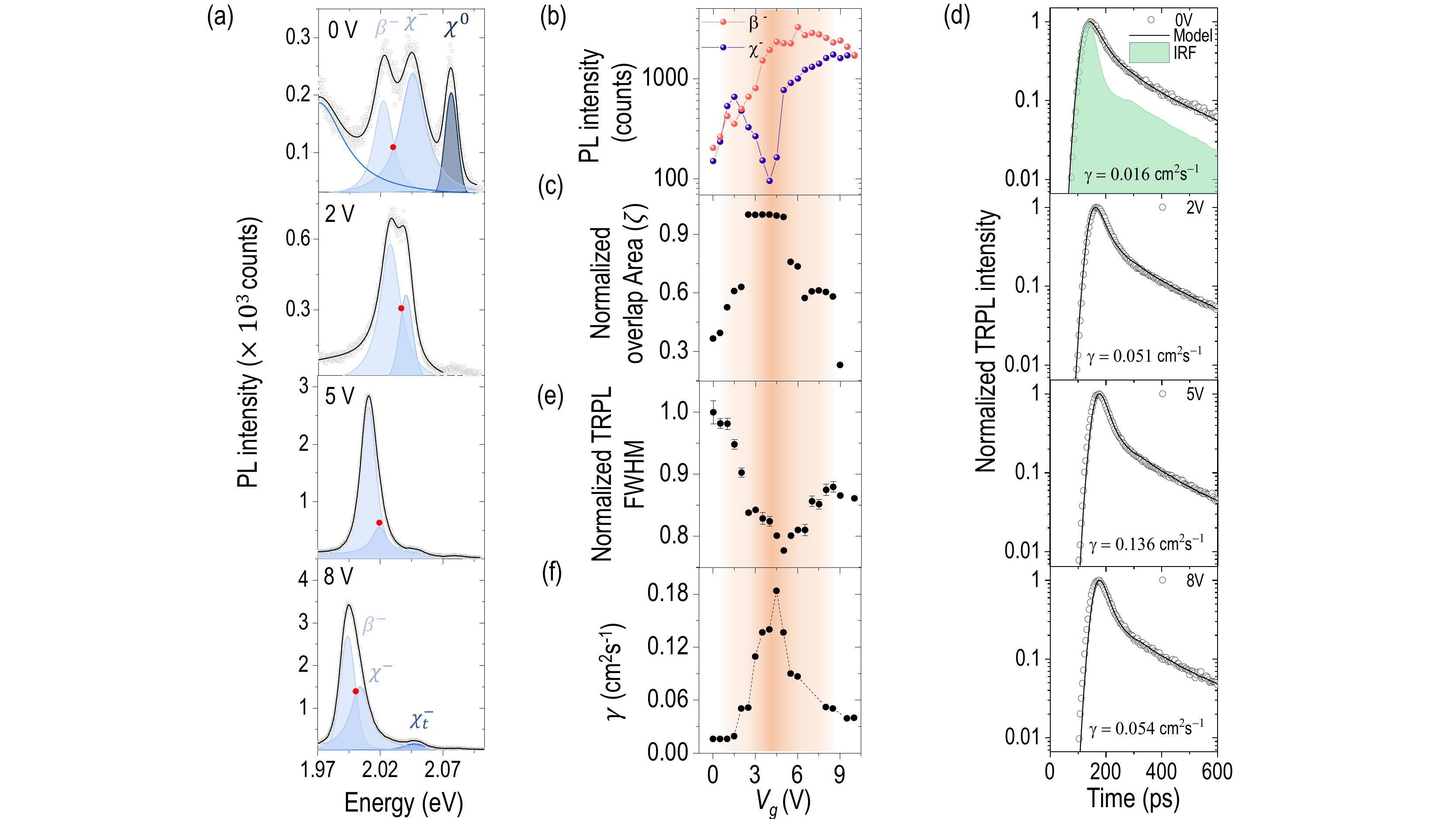}
	\vspace{-0.25in}
	\caption{\textbf{Evidence of Auger induced trion-trion annihilation.} (a) PL peak evolution for a few representative $V_g$ values. The PL spectra and the corresponding fittings of individual peaks are shown by solid lines and shaded areas, respectively. At $V_g = 5$ V, almost complete overlap between $\chi^-$ and $\beta^-$ is observed, coupled with a strong suppression of $\chi^-$ peak intensity. (b) PL intensity values of the fitted  $\chi^-$ (blue spheres) and $\beta^-$ (red spheres) peaks as a function of $V_g$. The shaded region indicates strong Auger regime. (c) Spectral area overlap of $\chi^-$ and $\beta^-$, normalized with respect to $\chi^-$ area, is plotted with increasing $V_g$. The striking anti-correlation with the $\chi^-$ peak intensity in (b) is conspicuous in the strong Auger region. (d) TRPL data (open symbols) is fitted with equation \ref{eq:TRPL} (solid traces) after deconvolution with the IRF, at representative $V_g$ values. IRF is shown by the light blue shaded area for $V_g = 0$ V. (e) Normalized TRPL full-width at half-maxima (FWHM) is plotted with respect $V_g$, indicating a non-monotonic trend with a reduced value in the strong Auger region. (f) Fitted values of the Auger coefficient ($\gamma$) from (d) are plotted with $V_g$, again showing non-monotonic variation. The deep shaded region covering (b), (c), (e) and (f) represents dominant Auger effect, which gradually reduces on both sides. } \label{fig:trpl and overlap}
\end{figure}

The PL spectra of some representative $V_g$ values, along with the fitting of individual peaks (shaded area), are shown in Figure \ref{fig:trpl and overlap}a. The variation of the PL intensity of the $\chi^-$ and $\beta^-$ peaks as a function of $V_g$ is shown in Figure \ref{fig:trpl and overlap}b. We plot the corresponding spectral overlap ($\zeta$) between the $\chi^-$ and $\beta^-$ peaks (calculated as the overlap area between the two peaks normalized by the area under the $\chi^-$ peak) in Figure \ref{fig:trpl and overlap}c (after spectral overlap, fitting with a single peak and total peak FWHM is shown in Supplementary Material \cite{Supp.Mater.} Figures 6 and 7, respectively). It is striking to note that the $\chi^-$ peak intensity is highly non-monotonic with $V_g$. In particular, after an initial increase ($V_g \leq 1.5$ V), there is an anomalous suppression in the $\chi^-$ peak intensity with an increase in $V_g$ in the range of $1.5$ V $\leq V_g \leq 5$ V (shaded region), beyond which it again increases strongly. Comparing Figures \ref{fig:trpl and overlap}b and c reveals that the $\chi^-$ peak intensity has strong negative correlation with $\zeta$.

Further, we observe no such anomalous decrease in trion intensity when we excite the same sample with a continuous wave laser (532 nm) at a power of 8.04 $\mu W$ (see Supplementary Material \cite{Supp.Mater.} Figure 8 for more details). The excitation density is $\approx 4\times10^3$-fold enhanced in the pulsed excitation mode compared with the continuous wave case.

\begin{figure}[!hbt]
	\centering
	\vs{-0.1in}
	\hs{-0.5in}
	\includegraphics[scale=0.5]{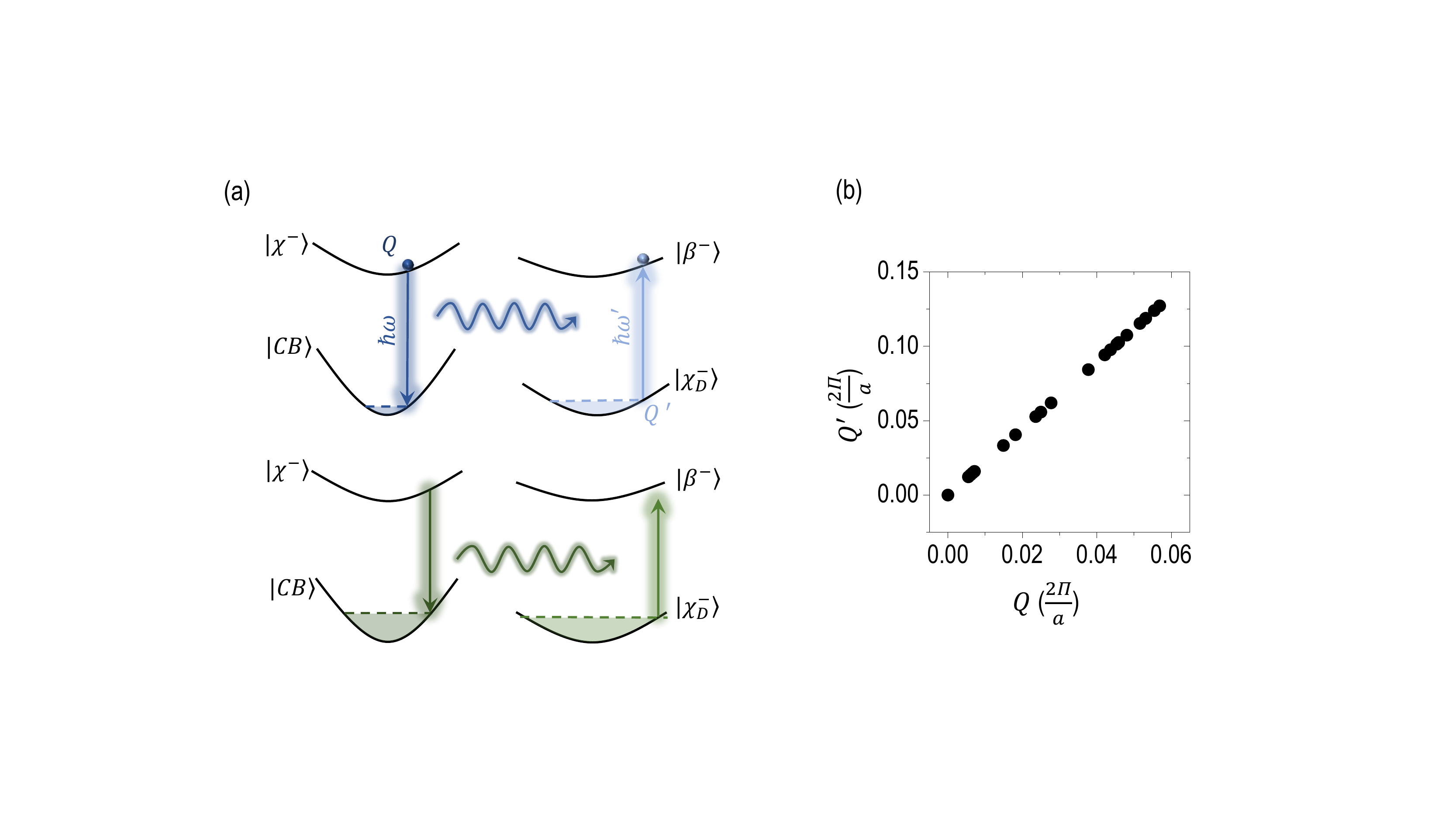}
	\vspace{-1in}
	\caption{\textbf{Schematic diagram depicting the resonant Auger mechanism.} (a) Schematic representation of the Auger interaction between two trions ($\chi^-$ and $\chi_D^-$). When the spectral overlap between $\beta^-$ and $\chi^-$ transition is strong, the non-radiative annihilation of a $\chi^-$ gives rise to the transition of one $\chi_D^-$ to $\beta^-$ state (top panel). When the spectral overlap is reduced, such strong resonance is eliminated, suppressing the Auger efficiency (bottom panel). (b) COM momentum for $\beta^-$ ($Q^\prime$) is plotted with respect to that of $\chi^-$ ($Q$), obtained from equation 3. The energy corresponds to the maximum overlap point, indicated by the red dots in Figure \ref{fig:trpl and overlap}a.}\label{fig:mechanism}
\end{figure}

These observations directly indicate the role of TTA in the anomalous suppression of the trion emission intensity. The mechanism is explained in Figure \ref{fig:mechanism}a (top panel). The measured photon energy $\hbar \omega$ ($\hbar \omega^\prime$) of the $\chi^-$ ($\beta^-$) emission peak corresponds to the transitions between the states $\ket {\chi^-}$ and $\ket {CB}$ ($\ket {\beta^-}$ and $\ket{\chi_D^-}$). When the spectral overlap is large, there is a strong resonance between these two transitions. Accordingly, at high excitation density, a bright trion (with COM $\hbar Q$) can undergo non-radiative recombination to the conduction band, providing its recombination energy to a dark trion (with COM momentum $\hbar Q^\prime$). This receiving dark trion, with this energy, moves to the $\beta^-$ state (with COM momentum $\hbar Q^\prime$) due to the presence of the strong resonance between these two transitions. The $\beta^-$ eventually emits a photon and relaxes back to the $\chi_D^-$ state. The net result of this entire process is the annihilation of a bright trion and the transfer of its oscillator strength for the formation of a charged biexciton. Such an Auger induced TTA process efficiently suppresses the bright trion emission intensity in the range of $1.5$ V $\leq V_g \leq 5$ V (shaded region in Figures \ref{fig:trpl and overlap}a-b). However, when the spectral overlap between $\beta^-$ and $\chi^-$ decreases ($V_g > 5$ V), the transition from $\ket{\chi_D^-}$ to $\ket{\beta^-}$ state is less efficient due to breaking of the resonance, as depicted in Figure \ref{fig:mechanism}a (lower panel).

Energy and linear momentum conservation during the Auger process provides the following relation between the momenta of the participating trion and biexciton:
\begin{equation}
{Q^\prime}^2(V_g) = -\frac{15m_0}{2\hbar^2}\delta(V_g=0) + 5Q^2(V_g)
\end{equation}
where $\delta = \hbar(\omega-\omega^\prime)$. We assume that the effective mass for electron and hole are 0.5$m_0$ each. Using the maximum overlap energy value (indicated by the red dots in Figure \ref{fig:trpl and overlap}a) as the representative points for Auger process, the relation between $Q$ and $Q^\prime$ (corresponding to different $V_g$) is depicted in Figure \ref{fig:mechanism}b.

As stated before, when the spectral resonance reduces ($V_g > 5$ V), TTA is gradually suppressed, and  $\chi^-$ peak intensity increases with $V_g$, as usual (Figure \ref{fig:trpl and overlap}b). We also note, after $V_g> 6$ V (when $\zeta$ is small), the $\beta^-$ peak intensity reduces with an increase in $V_g$, which could be due to a less generation of charged biexciton from the dark trion due to the suppressed Auger process.

To support the above-mentioned TTA mechanism, we perform \emph{in situ} TRPL measurement on the sample (see Methods section for experimental details). The raw time-resolved spectra are provided in Figure \ref{fig:trpl and overlap}d (open symbols) at representative $V_g$ values. In Figure \ref{fig:trpl and overlap}e, we plot the extracted (normalized) full-width at half-maximum (FWHM) of the decay plot, which indicates a faster decay in the regime where the trion intensity is suppressed. The observation of a faster decay of the trion at larger spectral overlap, coupled with reduced emission intensity, suggests an increasingly faster non-radiative process, supporting the TTA mechanism. At larger $V_g$ ($> 5$ V), when the spectral overlap is smaller, TTA is suppressed, leading to an increase in the FWHM of the decay plot (Figure \ref{fig:trpl and overlap}e).

The decay of the time dependent bright trion population density [$n(t)$] at $Q$ is given by
\begin{equation}
\frac{dn(t)}{dt} = -\frac{n(t)}{\tau} - \gamma n(t)n_d(t)
\end{equation}
Here, $\tau$ is the trion decay timescale which includes both radiative and non-radiative mechanisms (other than TTA), while $\gamma$ is the Auger coefficient for the TTA process. $n(t)$ and $n_d(t)$ are the bright and dark trion population density at $Q$ and $Q^\prime$, respectively. Note that both the bright and dark trion bands have similar curvature. Given that we are considering a two-dimensional system, the density of states is independent of the momentum. Thus, both the bands have a similar density of states irrespective of the values of $Q$ and $Q^\prime$. In addition, we also expect that at these momentum values, the filling factors are similar since trion and biexciton radiative recombination peaks around $Q$ and $Q^\prime$, respectively. Hence, we approximate $n_d(t) \approx n(t)$. Then $n(t)$ can be written as:
\begin{equation}\label{eq:TRPL}
n(t) = \frac{n(0) e^{-t/\tau}}{1+[n(0)\gamma\tau](1-e^{-t/\tau})}
\end{equation}
$n(0)$ is the trion population at time $t = 0$, and is estimated to be $\sim 10^{12}$ cm$^{-2}$ at the given laser power. \textcolor{black}{Finally, the experimental TRPL spectra is fitted with
\begin{equation}\label{eq:improved_TRPL}
n^\prime(t) = n(t)+\beta_1e^{-t/\tau_1}+\beta_2e^{-t/\tau_2}
\end{equation}
by deconvoluting with the instrument response function (IRF). The second and third terms respectively take into account the decay from the charged biexciton and small defect tail in the spectral overlapping region. We use $\gamma$, $\beta_1$ and $\beta_2$ as fitting parameters. The respective values of $\tau$, $\tau_1$ and $\tau_2$ are obtained from the non-TTA region (that is, $V_g<0$ V) and kept as constant at $\tau=14$ ps, $\tau_1=50$ ps, and $\tau_2=750$ ps (see Supplementary Material \cite{Supp.Mater.} Figures 9a and b for details). These obtained lifetime values for trion and charged biexciton agree well with existing reports \cite{wang2014valley,liu2019gate,nagler2018zeeman}. The obtained fittings of the experimental data in the TTA region with equation \ref{eq:improved_TRPL} are shown as the solid traces in Figure \ref{fig:trpl and overlap}d. The value of $\gamma$ ensuring the best fit with the experimental data is plotted in Figure \ref{fig:trpl and overlap}f at different gate voltages.}

It is clear from Figure \ref{fig:trpl and overlap}f that $\gamma$ is a strong function of $V_g$ induced spectral overlap between $\chi^-$ and $\beta^-$. In particular, we extract $\gamma=0.016$ cm$^2$s$^{-1}$ at small positive $V_g$, which increases by about ten-fold to $\gamma=0.18$ cm$^2$s$^{-1}$ at $V_g=4.5$ V. This suggests that the spectral overlap between the two transition energies (Figure \ref{fig:mechanism}a) is essential to enhance the TTA process.

In conclusion, we demonstrate a large tunability of charged exciton and charged biexciton spectral features by applying a gate voltage. This allows us to directly map the large center of mass momentum associated with these emitting species, in stark contrast to the zero-momentum neutral exciton. When the two charged species are close to spectral resonance, we observe an anomalous suppression of the bright trion intensity with increasing gate voltage, suggesting a trion-trion annihilation through the Auger process facilitated by the spectral resonance. We could tune the strength of the Auger process ten-fold by varying the gate-voltage-dependent spectral overlap. Such an electrically tunable resonant Auger process could be a prospective tool to explore more complex many-body phenomena among higher-order excitonic complexes.
\section*{Methods}
\textbf{Sample preparation.} The top-gated stack is prepared by mechanical exfoliation of the individual layers from bulk crystals of WS\tsub2, hBN, and graphite. First, the exfoliation is performed on Polydimethylsiloxane (PDMS) using 3M\textsuperscript{TM} scotch tape, and then the layers are transferred sequentially onto Si substrate covered with 285 nm thick SiO\tsub2. The few-layer graphene layers for the top gate electrode and the grounding electrode are connected to two different pre-patterned metal pads. The metal pads are fabricated using direct laser lithography, followed by sputtering of Ni/Au (10/50 nm) and lift-off. The deterministic transfer process is performed at room temperature under an optical microscope using a micromanipulator. After the transfer process is complete, the stack is annealed at 200$^{\circ}$C for 3 hours in a high vacuum chamber ($10^{-6}$ mbar) to ensure better adhesion between layers.\\
\textbf{Sample characterization.} A pulsed laser head of 531 nm wavelength (PicoQuant) operated by PDL-800D laser driver is used for optical excitation. The laser has a pulse width of 48 ps with a variable repetition rate (10 MHz is used in this work). A 50:50 beam splitter is used to divert half of the emission signal from the sample (see \textbf{Supporting Figure 1} for schematic) to a spectrometer for steady-state PL. The other half of the emission signal is directed to a single photon avalanche detector (SPD-050-CTC from Micro Photon Devices). The detector is connected to a Time-Correlated Single Photon Counting (TCSPC) system (PicoHarp 300 from PicoQuant). The time-resolved dynamics of the trion is obtained using a 610 nm bandpass filter (FWHM 10 nm) before the single-photon detector. The setup's instrument response function (IRF) has a decay of $\sim$ 23 ps and an FWHM of $\sim$ 52 ps.
\section*{Acknowledgements}
This work was supported in part by a Core Research Grant from the Science and Engineering Research Board (SERB) under Department of Science and Technology (DST), a grant from Indian Space Research Organization (ISRO), a grant from MHRD under STARS, and a grant from MHRD, MeitY and DST Nano Mission through NNetRA. Growth of hexagonal boron nitride crystals was supported by the Elemental Strategy Initiative conducted by the MEXT, Japan, Grant Number JPMXP0112101001, JSPS KAKENHI Grant Number JP20H00354 and JP19H05790.
\section*{Competing Interests}
The authors declare no competing financial or non-financial interests.
\section*{Data Availability}
Data available on reasonable request from the corresponding author.\\
%\section*{References}
\bibliographystyle{unsrt}
\bibliography{references}
\newpage
\newpage
\newpage
\newpage


\section*{Supplementary material}

# SUPPLEMENTARY MATERIAL

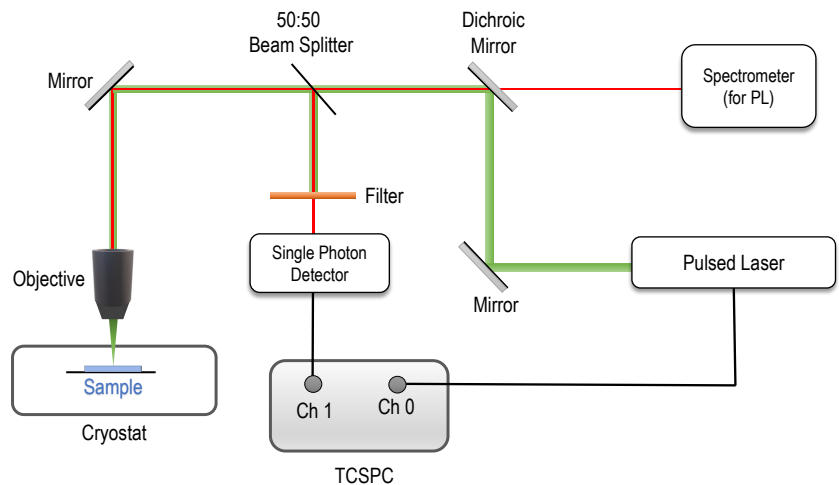

FIG. 1. **Schematic of the measurement setup for *in situ* TRPL and steady-state PL.** A pulsed laser source (pulse width ∼ 48 ps) of 531 nm wavelength and 10 MHz repetition rate is used for excitation. A temperature of 5 K is maintained in the cryostat during measurement.

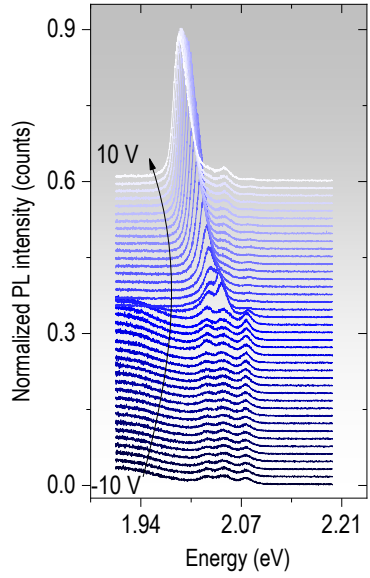

FIG. 2. **PL spectra at different gate voltages.** The PL spectra obtained at $V_g$ = -10 to 10 V. In the p-doping regime ($V_g < 0$ V) three PL peaks namely $\beta^+$, $\chi^0$ and $\chi^+$ (refer to main text Figure 1) do not show any sharp spectral shift, whereas in the n-doping regime ($V_g > 0$ V), $\chi^-$ and $\beta^-$ show strong redshift including a spectral overlap region.

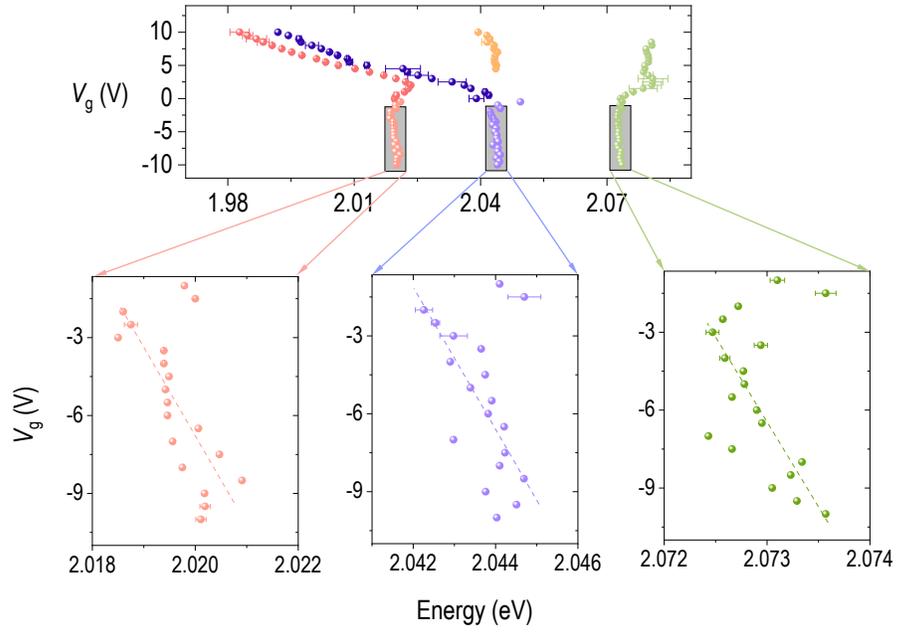

FIG. 3. **Blue shift of exciton, positive trion and positively charged biexciton.** We note about 2-4 meV blue shift of $\chi^0$, $\chi^+$ and $\beta^+$ (from left to right). The dashed lines are guide to the eye.

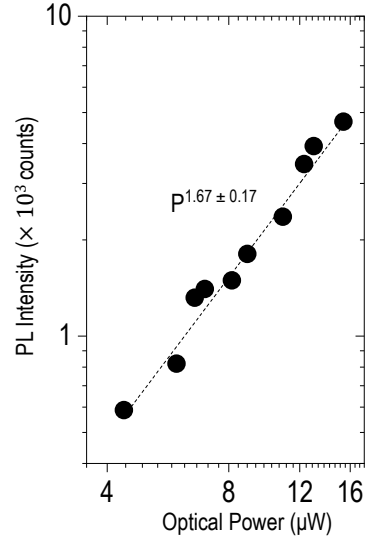

FIG. 4. **Power law fitting of charged biexciton with respect to laser power.** We fit the obtained charged biexciton PL intensity (at $V_g = 0$ V) with varying incident laser power. Fitting shows a super-linear relation of PL intensity of $\beta^-$ with laser power ($\propto P^{1.67\pm0.17}$).

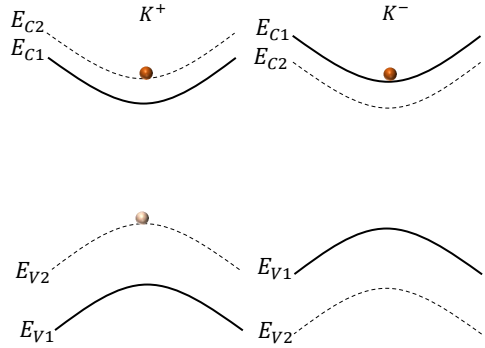

FIG. 5. **Schematic of intervalley trion configuration.** We observe an intervalley trion after $V_g > 4.5$ V, schematically depicted here. The extra electron along with the spin allowed bright electron-hole pair is taken from the top conduction band of the other valley.

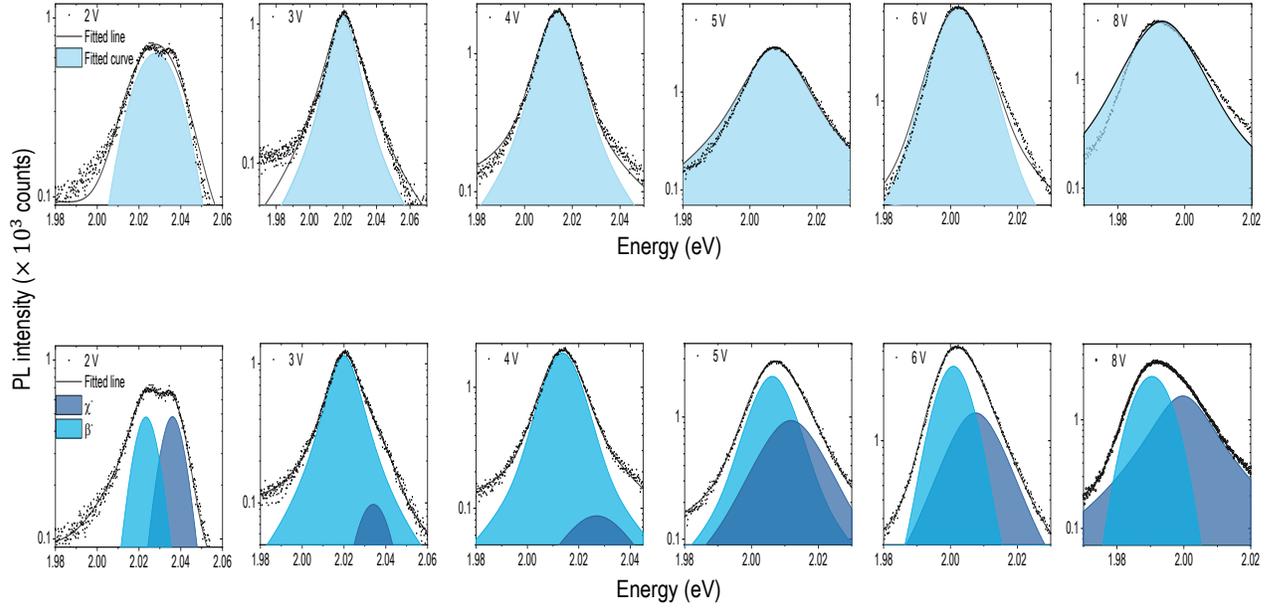

FIG. 6. **PL fitting comparison for the trion and charged biexciton merged peak with a single and double voigt curve.** We fit the $\chi^-$ and $\beta^-$ merged PL peak (in the $2 \leq V_g \leq 8$ V range) with a single as well as two voigt curves. Fitting with two voigt curves improve the fit quality significantly, proving existence of both $\chi^-$ and $\beta^-$ even after strong spectral overlap.

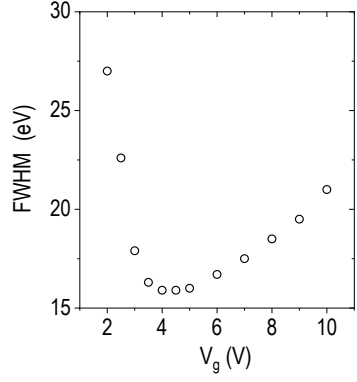

FIG. 7. **FWHM of the merged PL peak versus gate voltage.** The FWHM of the merged PL peak of $\chi^-$ and $\beta^-$ together is plotted with $V_g$. It shows a non-monotonic behavior with the minimum FWHM obtained when the overlap between trion and charged biexciton is maximum.

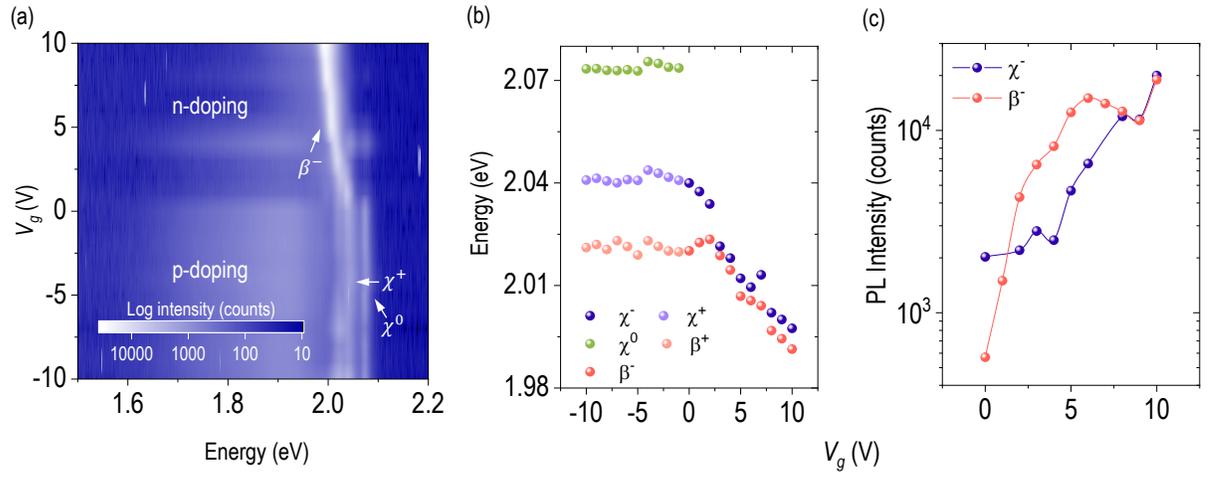

FIG. 8. **Continuous wave PL fitted peak position and intensity.** (a) PL data obtained using a 532 nm continuous wave laser is shown in a color plot. (b) Fitted peak positions showing spectral overlap between $\chi^-$ and $\beta^-$. (c) Fitted peak intensity values showing no appreciable reduction of $\chi^-$ PL intensity, even in the spectral overlap region.

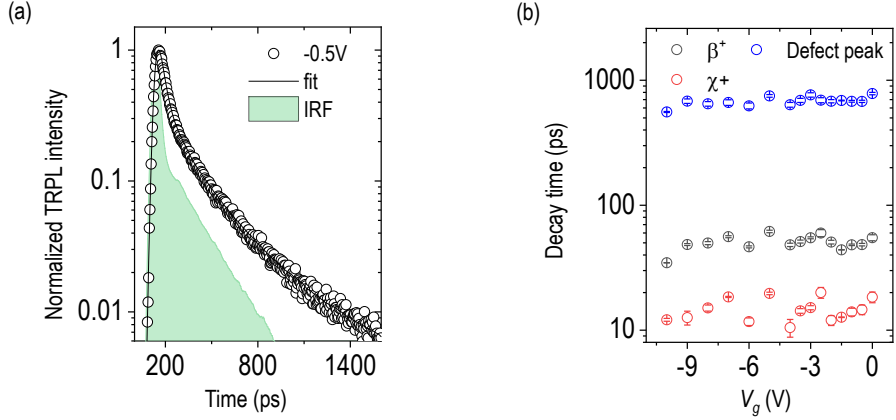

FIG. 9. **Extraction of $\tau$, $\tau_1$, and $\tau_2$.** To extract these parameters, we choose the voltage range where Auger-effect is negligible by ensuring minimal spectral overlap between trion and charged biexciton, that is, in the voltage range $-10 \leq V_g \leq 0$ V. In this region, since Auger effect is negligible, we fit the experimental TRPL data using the expression $Ae^{-t/\tau} + A_1 e^{-t/\tau_1} + A_2 e^{-t/\tau_2}$ (after deconvolution with the IRF) accommodating the decay for the trion, charged biexciton, and any small defect luminescence tail. (a) An example fit (in solid black trace) with the experimental spectra (in symbols) at $V_g = -0.5$ V. (b) The extracted value of $\tau$, $\tau_1$ and $\tau_2$ in the non-TTA range ($-10 \leq V_g \leq 0$ V), suggesting the values are independent of $V_g$. The used vales for $\tau$, $\tau_1$ and $\tau_2$ in the TTA regime are 14, 50, and 750 ps, respectively.


\end{document}